# Acceleration of Relativistic Particles in Counterpropagating Circularly Polarized Alfvén Waves

S. Isayama[1,2] , K. Takahashi[3], S. Matsukiyo[1,2] , and T. Sano[4]
[1] Department of Advanced Environmental Science and Engineering, Kyushu University, 6-1 Kasuga-Kohen, Kasuga, Fukuoka 816-8580, Japan
isayama@esst.kyushu-u.ac.jp
[2] International Space Center for Space and Environmental Science, Kyushu University, 744 Motooka, Nishi-Ku, Fukuoka 819-0395, Japan
[3] Interdisciplinary Graduate School of Engineering Sciences, Kyushu University, 6-1 Kasuga-Kohen, Kasuga, Fukuoka 816-8580, Japan
[4] Institute of Laser Engineering, Osaka University, Suita, Osaka 565-0871, Japan

## Abstract

Counterpropagating Alfvén waves are ubiquitously observed in many astrophysical environments, such as a star surface and a planetary foreshock. We discuss an efficient particle acceleration mechanism in two counterpropagating circularly polarized Alfvén waves. Phase transitions of particle behavior occur when wave amplitudes exceed two critical values. Above the critical amplitudes, the numerical simulation shows that any particles irreversibly gain relativistic energy within a short time regardless of their initial position and energy once the coherent waveform is formed. The accelerated particles have spatial coherence. Higher wave phase velocity requires smaller critical amplitudes, while the maximum attainable energy increases as the wavenumber and the frequency decrease. The results may be applicable in some astrophysical phenomena, as well as a future laboratory experiment using high-power lasers.

*Unified Astronomy Thesaurus concepts:* Heliosphere (711); Cosmic rays (329); Magnetars (992); Solar wind (1534)

## 1. Introduction

The acceleration mechanism of cosmic rays (CRs) has been unresolved since it was found by Hess in 1912 (Hess 1912). Coherent large-amplitude electromagnetic waves in a plasma should play crucial roles in the acceleration of high-energy CRs. In a terrestrial foreshock, a short large-amplitude magnetic structure (Giacalone et al. 1993) may accelerate a fraction of particles directly, or indirectly playing the role of a scatterer in the so-called diffusive shock acceleration process (Blandford & Eichler 1987). The large-amplitude Alfvén waves are often observed in the solar wind. For example, in the fast streams from coronal holes they are considered to play an important role in the generation of turbulence, which heats the corona and accelerates the solar wind (Goldstein et al. 1995).

In high-energy astrophysical environments, the quasiperiodic oscillations of the emissions from soft gamma-ray repeaters could be related to Alfvén waves excited near the surface of a neutron star (Blaes et al. 1989; Israel et al. 2005; Wang 2020). An energetic starquake drives Alfvén waves in the magnetosphere of the source object, and part of the dissipated energy of the waves accelerates particles (Blaes et al. 1989; Holcomb & Tajima 1991). Chen et al. (2002) proposed the idea of Alfvénic wakefield acceleration near a relativistic shock. The intense Alfvén waves are thought to be excited also in an accretion disk of a black hole. Ebisuzaki & Tajima (2014, 2021) discussed the scenario in which the Alfvén waves are converted to the light waves as they propagate in the jet along the field line in a rarefied plasma and particles are energized through the wakefield acceleration mechanism (Ligorini et al. 2021).

In the above wide variety of examples, coherent large-amplitude wave–particle interactions are inherently assumed. Large-amplitude waves spontaneously generate phase coherence among a finite number of wave modes (Nariyuki & Hada 2006). The generation of a wakefield is a typical example of such a process, which is also known as a parametric instability. In the lowest order, a parametric instability occurs through three-wave interactions. The decay instability of an Alfvén wave is one of the common examples. In this process two Alfvén waves propagating parallel and antiparallel to the ambient magnetic field couple with a longitudinal wave. It has been extensively discussed that the longitudinal wave causes plasma heating contributing to the acceleration of the solar corona (Shoda et al. 2019). Furthermore, multiple transverse (or Alfvén) waves generated through successive decay instabilities lead to turbulent acceleration of particles (Comisso & Sironi 2018).

More than a decade ago, Matsukiyo & Hada (2009) showed that a relativistic Alfvén wave in a pair plasma is unstable to form the coherent waveform of a magnetic envelope, which results in the particle acceleration being much more efficient than in a nonrelativistic case. The process requires the presence of two locally enhanced counterpropagating Alfvén waves. The particles pre-accelerated in the course of preceding successive decay instabilities are efficiently accelerated near a coherent magnetic envelope trough. However, the applicability of this acceleration process looks limited because the process requires pre-acceleration of particles. On the other hand, in extremely high energy situations such as in the examples mentioned in the previous paragraph, the generation of counterpropagating Alfvén waves may be easily realized because of their sufficiently large free energy. In terms of the decay instability, the growth rate is proportional to the wave amplitude. Even without the instability, in a region like a star surface where a closed magnetic loop is formed, an Alfvén wave driven at a footpoint may be reflected at a conjugate footpoint so that counterpropagating Alfvén waves may be naturally generated.

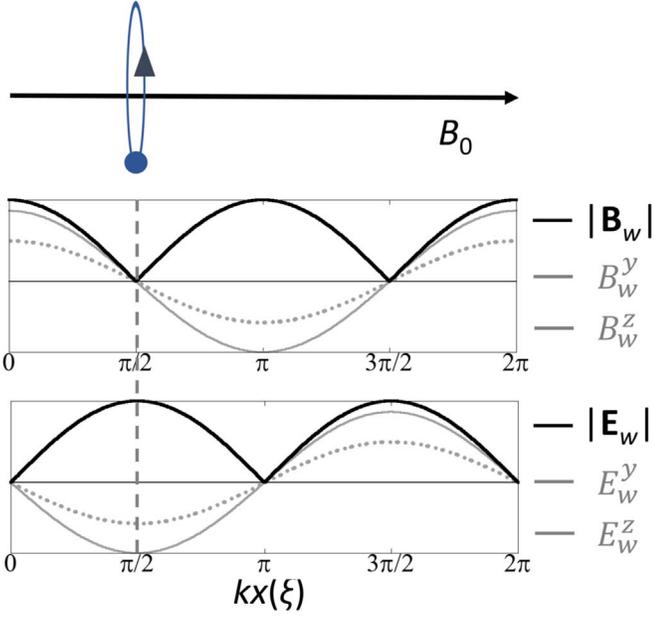

**Figure 1.** Typical waveforms of magnetic and electric fields of two counterpropagating waves described by Equations (1) and (2). In each panel the solid and dashed gray lines denote carrier waves at $\omega t = \pi/6$ and the solid black line shows the envelope.

In this work, we focus on the motion of charged particles in a system containing two circularly polarized, sufficiently large amplitude Alfvén waves with strong ambient magnetic field. We discuss the acceleration process in a Hamiltonian system by reducing the system's degrees of freedom and show that there exist critical wave amplitudes above which any particles are accelerated to the relativistic energy regardless of their initial energy. It will also be shown that the mechanism works for both electrons and ions and that accelerated particles have strong spatial coherence as well, which may be suitable for strong radiation from the objects. We begin with a review of particle motion in the presence of two counterpropagating circularly polarized waves (Section 2.1). Then, we discuss the phase transitions of particle orbits when wave amplitudes exceed two critical wave amplitudes (Section 2.2). Test particle simulation (Section 3.1) and particle-in-cell (PIC) simulation (Section 3.2) are performed to confirm the particle acceleration process during the Alfvén wave collision.

## 2. Theory

### 2.1. Basic Equations

We discuss particle motion in the presence of two counterpropagating circularly polarized waves. The same amplitude, frequency, and wavenumber are used for these two waves. The magnetic and electric fields of the combined wave are given as

$$\boldsymbol{B}_w \equiv \begin{pmatrix} B_w^y \\ B_w^z \end{pmatrix} = 2B_w \cos kx \begin{pmatrix} \cos \omega t \\ \sin \omega t \end{pmatrix} \quad (1)$$

$$\boldsymbol{E}_w \equiv \begin{pmatrix} E_w^y \\ E_w^z \end{pmatrix} = -\frac{2\omega}{k} B_w \sin kx \begin{pmatrix} \cos \omega t \\ \sin \omega t \end{pmatrix}. \quad (2)$$

Here $k$ and $\omega$ denote the wavenumber and the frequency, respectively. Hereafter, we assume that $\omega$ is positive for right-hand polarized (with respect to an ambient magnetic field) (R) waves and negative for left-hand polarized (L) waves and that $k > 0$. Figure 1 shows the modeled wave magnetic and electric fields. It is clear from Equations (1) and (2) that their envelopes are independent of time. We consider the motion of a charged particle in the above electromagnetic fields. The equation of motion is

$$\frac{d\boldsymbol{u}}{dt} = \frac{q_s}{m_s}\left[\boldsymbol{E}_w + \frac{\boldsymbol{u}}{\gamma} \times (\hat{\boldsymbol{x}} B_0 + \boldsymbol{B}_w)\right], \quad (3)$$

where $\boldsymbol{u}$ denotes the four-velocity, $\gamma = \sqrt{1 + u^2}$, $q_s$ and $m_s$ indicate the charge and the mass of the $s$th species, and $B_0$ is the strength of the ambient magnetic field along the $x$-axis. By introducing $\boldsymbol{u} = (u_\parallel(t), u_\perp(t)\cos\phi(t), u_\perp(t)\sin\phi(t))$, Equation (3) is written as follows:

$$\dot{u}_\parallel = -2\epsilon_s b_w \frac{u_\perp}{\gamma} \cos\kappa\xi \sin\psi \quad (4)$$

$$\dot{u}_\perp = -2\epsilon_s b_w \left( v_{\rm ph} \sin\kappa\xi \cos\psi - \frac{u_\parallel}{\gamma} \cos\kappa\xi \sin\psi \right) \quad (5)$$

$$\dot{\psi} = 2\epsilon_s \frac{b_w}{u_\perp}\left( v_{\rm ph} \sin\kappa\xi \sin\psi + \frac{u_\parallel}{\gamma}\cos\kappa\xi \cos\psi \right)$$
$$- \left( \nu + \epsilon_s \frac{1}{\gamma} \right). \quad (6)$$

Here we used the normalized variables, $\xi = x\Omega_{cs}/c$, $\tau = \Omega_{cs} t$, $\kappa = kc/\Omega_{cs}$, $\nu = \omega/\Omega_{cs}$, $b_w = B_w/B_0$, and $v_{\rm ph} = \nu/\kappa$, where $\Omega_{cs} = |q_s|B_0/m_s$ is a nonrelativistic gyrofrequency of the $s$th species. Moreover, $\psi = \phi - \nu\tau$, and $\epsilon_s(=\pm 1)$ denotes the sign of a particle charge. The above equations are a generalized version of those discussed by Matsukiyo & Hada (2009).

To discuss the particle acceleration by reducing the degree of freedom, we focus on the motion of a charged particle that is at a trough (crest) of the magnetic (electric) envelope by fixing $\kappa\xi = \pi/2$ and $u_\parallel = 0$ ($\dot{\xi} = u_\parallel/\gamma$) (later in the test particle simulation, we will show that all the nonrelativistic particles are first accumulated at this position by the Lorenz force and accelerated to relativistic energy even when the parallel fluctuation is included). Then, Equations (4)–(6) are reduced to the following two equations:

$$\dot{p} = -4\epsilon_s \sqrt{p}\, b_w v_{\rm ph} \cos\psi \quad (7)$$

$$\dot{\psi} = 2\epsilon_s \frac{b_w}{\sqrt{p}} v_{\rm ph} \sin\psi - \left( \nu + \epsilon_s \frac{1}{\sqrt{1+p}} \right). \quad (8)$$

Here $p = u_\perp^2$ has been introduced. This system has a constant of motion defined as

$$H(p,\psi) = -4\epsilon_s \sqrt{p}\, b_w v_{\rm ph} \sin\psi + \nu p + 2\epsilon_s\sqrt{1+p}, \quad (9)$$

where $\dot{p} = \partial H/\partial \psi$ and $\dot{\psi} = -\partial H/\partial p$ are satisfied. In the following we discuss the motion of an ion ($\epsilon_s = 1$) in two left-hand circularly polarized waves ($\nu < 0$). (Essentially the same phenomenon occurs also for an electron ($\epsilon_s = -1$) in two right-hand circularly polarized waves ($\nu > 0$)).

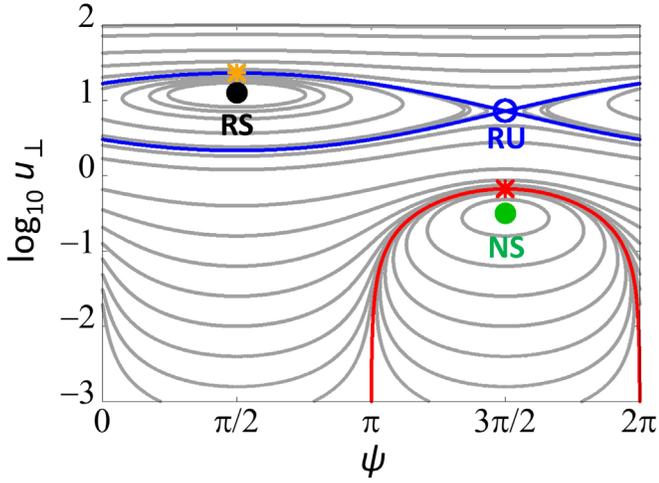

**Figure 2.** Trajectories of ions (gray lines) trapped at $\kappa \xi = \pi/2$ ($u_\parallel = 0$) in $\psi$–$u_\perp$ space for the case of $b_w = 0.15$, $v_{\rm ph} = -0.9$, and $\nu = -0.1$. Fixed points are indicated by circles (see details in the text). The relativistic and nonrelativistic trapping regions are bounded by the blue and red separatrices, respectively. The maximum energy point in the relativistic (nonrelativistic) trapping region is denoted by the orange (red) asterisk.

**Table 1**
Fixed Points at $\kappa \xi = \pi/2$ and $u_\parallel = 0$

| Type | $\psi$ | $p$ |
|---|---|---|
| RS | $\frac{\pi}{2}$ | $\frac{2b_w v_{\rm ph}}{\sqrt{p}} = \left(\nu + \frac{1}{\gamma}\right)$ |
| NS, RU | $\frac{3\pi}{2}$ | $-\frac{2b_w v_{\rm ph}}{\sqrt{p}} = \left(\nu + \frac{1}{\gamma}\right) \cdots$ (10) |

### 2.2. Phase Transition

A number of ion trajectories in $\psi$–$u_\perp$ space for different values of $H$ are shown by the gray lines in Figure 2, assuming $v_{\rm ph} = -0.9$, $\nu = -0.1$, and $b_w = 0.15$. There are two trapping regions inside the blue and red separatrices. The upper one (blue) exists only when the relativistic effect is taken into account ($\gamma \neq 1$) as discussed by Matsukiyo & Hada (2009). The corresponding stable and unstable fixed points are denoted as the relativistic stable point (RS) and the relativistic unstable point (RU), respectively. At the RS, a particle simultaneously resonates with the two waves. The lower one (red), on the other hand, exists even in nonrelativistic case ($\gamma = 1$) only if wave amplitude is finite. The corresponding stable fixed point is indicated as the nonrelativistic stable point (NS). The above fixed points are summarized in Table 1. The maximum momentum of the relativistic trapping (orange asterisk) in the case of $\gamma \gg 1$ is given by

$$u_{\perp,\max} \sim \left(\sqrt{\frac{2b_w}{\kappa}} + \sqrt{\frac{1}{|\nu|}}\right)^2. \quad (11)$$

Figures 3(b) and (c) show how particle behavior changes as the wave amplitude, $b_w$, increases. When $b_w = b_{w,1} \approx 0.33$ (Figure 3(b)), the RU (blue open circle) and the maximum energy point of the nonrelativistic trapping region (red asterisk) merge. This critical amplitude, $b_{w,1}$, is given by solving Equations (9) and (10) with $H(p_{\rm I}, 3\pi/2) = 2$, where $p_{\rm I}$ is the solution of Equation (10). In the limit of $p \gg 1$,

$$\sqrt{p_{\rm I}} \sim -\frac{2b_w v_{\rm ph} + 1}{\nu} \quad (12)$$

and

$$b_{w,1} \sim \frac{-1 + \sqrt{-2\nu}}{2v_{\rm ph}}, \quad (13)$$

where $b_w v_{\rm ph} > -1/2$. With this critical wave amplitude, a particle initially having nonrelativistic momentum and $\pi < \psi < 2\pi$ is accelerated to $p \sim p_{\rm I}$, and a particle initially having nonrelativistic momentum and $0 < \psi < \pi$ is accelerated to $p \sim u_{\perp\max}^2$. When the wave amplitude further increases to $b_w = b_{w,2} \approx 0.39$, the RU and the NS merge so that the trapping region around the NS disappears as shown in Figures 3(a) and (c). The corresponding value of $p = p_{\rm II}$ is given as a multiple root of Equation (10). In the limit of $p^2 \gg 1$,

$$\sqrt{p_{\rm II}} \sim -\frac{2(2b_w v_{\rm ph} + 1)}{3\nu} \quad (14)$$

and

$$b_{w,2} \sim \frac{3|\nu|^{2/3} - 2}{4v_{\rm ph}}. \quad (15)$$

In this situation all the nonrelativistic ions can be accelerated to $p \sim u_{\perp\max}^2$. It is noted from Equations (13) and (15) that a large $v_{\rm ph}$ results in small critical amplitudes.

## 3. Numerical Simulation

### 3.1. Test Particle Simulation

To discuss the effect of the abovementioned phase transition in the case without the restriction that $k\xi = \pi/2$ and $u_\parallel = 0$, we performed test particle simulations. Initially 10,000 particles are randomly located in $0 < \kappa \xi < 2\pi$ and their momentum distribution function is given by a Maxwellian, $f(u_{x,y,z}) = \exp(-u^2/2u_{\rm th}^2)/(2\pi)^{3/2} u_{\rm th}^3$, where $u_{\rm th} = 10^{-2}$. This initial distribution is shown as the gray line in Figure 4(a) and as the gray lines in Figure 4(b). For all these particles, Equation (3) is numerically solved by using a standard Buneman–Boris algorithm with the time step of $\Delta \tau = 0.04$. Equations (1) and (2) are used for the wave electromagnetic fields. Three runs were conducted, with (i) $b_w = 0.15 (< b_{w,1})$, (ii) $b_w = 0.35 (> b_{w,1})$, and (iii) $b_w = 0.4 (> b_{w,2})$, where $v_{\rm ph} = -0.9$ and $\nu = -0.1$ were fixed.

In the two panels the distributions at $\tau = 1600$ are plotted with different colors, i.e., (i) black, (ii) red, and (iii) blue, respectively. In run (i) the maximum perpendicular momentum ($\log_{10} u_\perp \sim 0.6$) roughly coincides with the value at the maximum energy point of the nonrelativistic trapping region (red asterisk) in Figure 2. In runs (ii) and (iii), on the other hand, all the particles were accelerated to relativistic energy and their maximum perpendicular momentum is $\sim u_{\perp\max}$. Figure 4(c) shows the particle positions as a function of time. All the particles first move toward the magnetic envelope trough, $\kappa \xi = \pi/2$ and $3\pi/2$. This is caused by the $\boldsymbol{u}_\perp \times \boldsymbol{B}_w$ force acting on each particle, which is also called the ponderomotive force. Once they approach the envelope trough, they are accelerated. Therefore, the accelerated particles have spatial coherence. Note that the acceleration occurs irreversibly. The reason for this is explained by the stability of the particle orbits in $u_\parallel(u_x)$. As discussed by Matsukiyo & Hada (2009), the system becomes unstable for parallel disturbances, $\delta u_\parallel$, around $\kappa\xi = \pi/2$ $(3\pi/2)$ when $0 < \psi < \pi$ $(\pi < \psi < 2\pi)$, while the system becomes stable around $\kappa\xi = \pi/2$ $(3\pi/2)$ when $\pi < \psi < 2\pi$ $(0 < \psi < \pi)$. The particles are initially accumulated at a trough (crest) of the magnetic field (electric field) $(\kappa\xi = \pi/2$ $(3\pi/2))$ and accelerated moving along the equi-$H$ lines in $u_\perp$–$\psi$ space. When they move in the region of $0 < \psi < \pi$, they are pushed away from the envelope trough so that their orbits deviate from the gray lines. The deviation is negligible when $u_\perp \ll 1$, since the right-hand side of Equation (4) is small. In other words, the deviation is significant when the particle momentum becomes relativistic. The corresponding orbits in runs (ii) and (iii) are denoted in Figures 4(d) and (e), respectively. Once detrapped, they show almost free orbits with a roughly constant $u_\perp$ close to the value given by Equation (11), $u_{\perp,\text{max}} \sim 34$. Figure 4(f) shows the time evolution of particle energy for runs (i)–(iii). All of the particles gain relativistic energy, $(\gamma-1) \gtrsim 1$, by $\tau \sim 10$. Furthermore, they reach the maximum energy, $\gamma \sim 10^2$, within $\tau \sim$ a few $\times\ 10^2$, corresponding to the gyroperiod of the corresponding relativistic particle. The values of $\gamma$ and $u_{\perp,\text{max}}$ suggest that particles are accelerated also in the parallel direction. This is caused by the Lorentz force, $\boldsymbol{u_\perp} \times \boldsymbol{B_w}$. We emphasize that this acceleration time is extremely short compared with any known stochastic acceleration processes, such as the second- and first-order Fermi acceleration. The results shown above are essentially independent of initial particle distribution function. For instance, we observed almost the same final particle distributions as in Figures 4(a) and (b) when we use the initial distribution function with oscillating bulk velocity satisfying the Walen relation (Equation (18) below). We also confirmed that this acceleration mechanism works even when the amplitudes, frequencies, and wavenumbers are not the same for the two counterpropagating waves (not shown). Further, it was verified that the acceleration works when particles initially have finite $u_\parallel$ satisfying the resonance condition with only one of the two waves. Figure 5 shows the result when the parallel initial velocity distribution is given as a shifted Maxwellian $f(u_x) = \exp(-(u-u_0)^2/2u_{\rm th}^2)/(2\pi)^{1/2}u_{\rm th}$ ($f(u_{y,z})$ are same as in Figure 4) with $u_0 = -4.69$, which satisfies the relativistic resonance condition of $\gamma\nu \sim -(1+\kappa u_\parallel)$ with the forward propagating wave. One should note that the presence of a well-defined trough of the magnetic envelope is necessary for this acceleration mechanism to work. This may require that the two Alfvén waves have enough amplitude, although they can have different amplitudes. If one of the two waves has too small an amplitude, a classical single wave–particle interaction (Bellan 2013) becomes dominant.

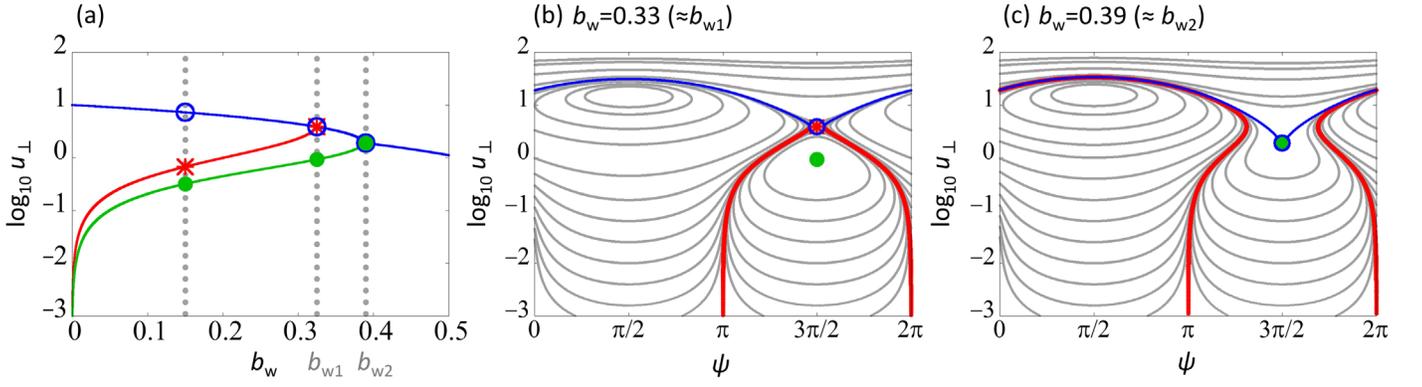

**Figure 3.** (a) Behaviors of the fixed points (NS, RU) and the maximum energy point of the nonrelativistic trapping region as a function of wave amplitude for the case of $v_{ph} = -0.9$ and $\nu = -0.1$. Phase transition occurs at the two critical amplitudes $b_w = b_{w,1}$ and $b_w = b_{w,2}$. (b) Ion trajectories (gray lines), separatrices, and some characteristic points in $u_\perp$–$\psi$ space for $b_w = 0.33 (\approx b_{w,1})$. (c) Same as panel (b), but for $b_w = 0.39 (\approx b_{w,2})$.

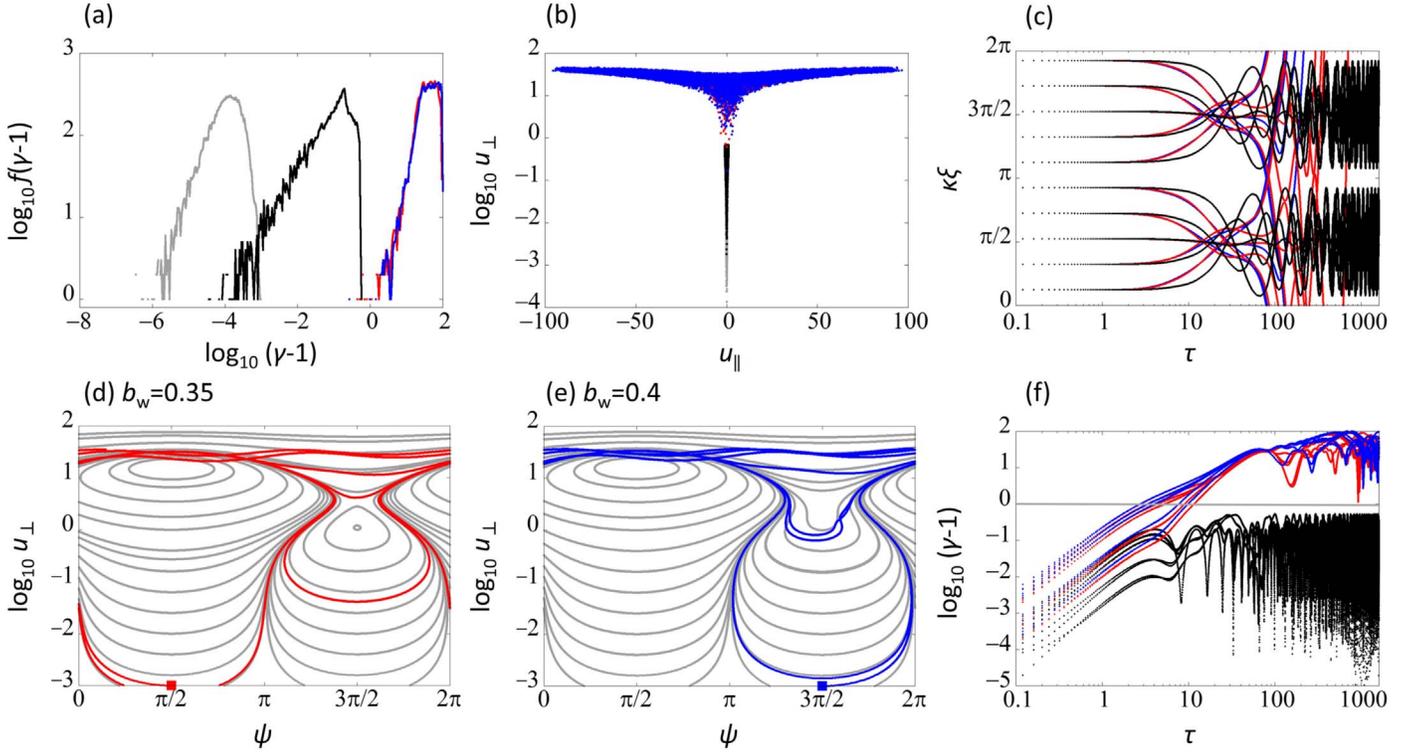

**Figure 4.** Results of test particle simulations for runs with (i) $b_w = 0.15$ (black), (ii) $b_w = 0.35$ (red), and (iii) $b_w = 0.4$ (blue). (a) Energy distribution of ions at $\tau = 1600$ (gray line indicates the initial distribution). (b) Ion distribution in $u_\perp$–$u_\parallel$ space at $\tau = 1600$ (gray lines denote the initial distribution). (c) Time histories of the position of 10 ions equally spaced initially. (d, e) Examples of ion trajectory in $u_\perp$–$\psi$ space for runs (ii) and (iii), where the initial points are denoted by squares. (f) Energy time histories of the 10 ions shown in panel (c).

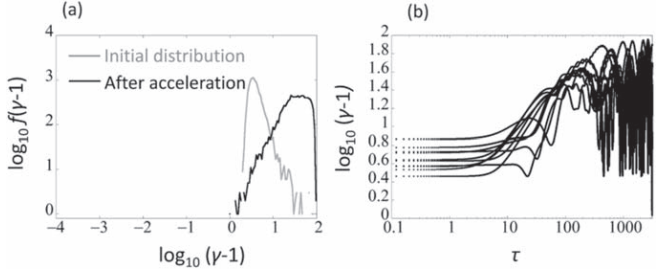

**Figure 5.** Results when ions initially have finite $u_\parallel$ satisfying the resonance condition with the forward propagating wave. (a) Energy distribution of ions at $\tau = 1600$ (gray line indicates the initial distribution). (b) Energy time histories of the 10 ions. Here $\nu = -0.1$, $v_{\rm ph} = -0.9$, and $b_w = 0.4$ are used for two counterpropagating waves.

### 3.2. PIC Simulation

A one-dimensional PIC simulation is performed to investigate the acceleration process of charged particles in a self-consistent system. The initial condition is shown in Figure 6. A monochromatic left-hand circularly polarized Alfvén wave packet propagating with $v_{\rm ph} > 0$ ($<0$) is initially given in $2150 < \xi < 2750$ ($2920 < \xi < 3520$). The edges of these wave packets are smoothed by adding the linear profiles of the width $\lambda = 2\pi/\kappa \sim 60c/\omega_{pe}$. The electromagnetic fields and the bulk velocities of electrons and ions in the packets are given as $B_y + iB_z = b_w \exp i(\kappa\xi - \nu\tau)$, $E_y + iE_z = e_w \exp i(\kappa\xi - \nu\tau)$, $u_{ey} + iu_{ez} = u_{e\perp} \exp i(\kappa\xi - \nu\tau)$, and $u_{iy} + iu_{iz} = u_{i\perp} \exp i(\kappa\xi - \nu\tau)$, respectively. This is a standard way to provide a circularly polarized wave (e.g., Terasawa et al. 1986). The amplitudes of each oscillation are given by

$$e_w = -i\kappa b_w/2\nu, \quad (16)$$

$$u_{e\perp} = -m_r e_w/(\nu\gamma_e + \Omega'_{ce}), \quad (17)$$

$$u_{i\perp} = e_w/(\nu\gamma_i - 1). \quad (18)$$

Here Equations (17) and (18) are known as the Walen relation (Hollweg et al. 1993). Parameters $\nu$, $\kappa$, $\gamma_e$, and $\gamma_i$ should satisfy

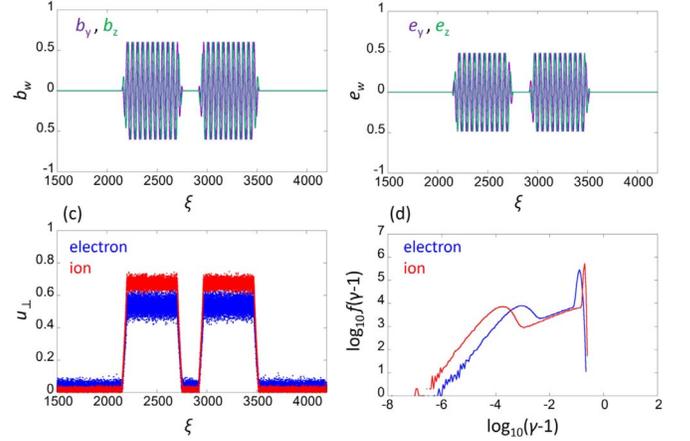

**Figure 6.** Initial profiles of (a) $b_y$ (purple) and $b_z$ (green), (b) $e_y$ (purple) and $e_z$ (green), and (c) $u_\perp$ of electrons and ions. (d) Initial energy distributions of electrons and ions. Hereafter, all quantities are evaluated at $1500 < \xi < 4170$ and the information of electrons and ions is shown in blue and red, respectively.

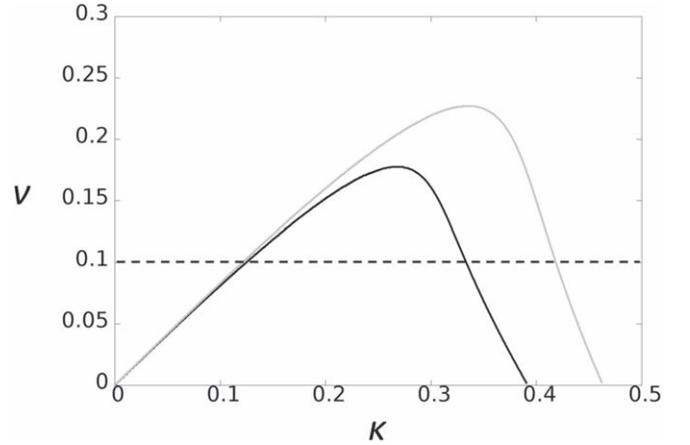

**Figure 7.** The dispersion relation of the Alfvén wave for the case of $b_w = 0.3$ (gray) and $b_w = 0.6$ (black). The dotted black line indicates $\nu = 0.1$.

**Table 2**
Simulation Conditions

| Case | $b_w$ | $\kappa$ | $\gamma_e$ | $\gamma_i$ | $b_{w,1}$ | $b_{w,2}$ |
|---|---|---|---|---|---|---|
| Subcritical | 0.3 | 0.121 | 1.031 | 1.159 | 0.33 | 0.41 |
| Supercritical | 0.6 | 0.124 | 1.134 | 1.372 | 0.34 | 0.42 |

the dispersion relation (Figure 7; Matsukiyo & Hada 2003)

$$\nu^2 - \kappa^2 = \nu\omega'^2_{pe}[1/(\nu\gamma_i - 1) + m_r/(\nu\gamma_e + \Omega'_{ce})], \quad (19)$$

where the normalized variables of $\Omega'_{ce} = \Omega_{ce}/\Omega_{ci}$, $\omega'_{pe} = \omega_{pe}/\Omega_{ci}$, and $m_r = m_i/m_e$ are used. Our simulation is performed with $\nu = 0.1$, $\Omega'_{ce} = m_r = 5.0$, and $\omega'_{pe} = 1.24$ for the cases of $b_w = 0.1$–0.8. Table 2 shows the simulation conditions of the two representative cases when the wave amplitude is subcritical ($b_w = 0.3 < b_{w,1}$) and supercritical ($b_w = 0.6 > b_{w,2}$). Figure 8 corresponds to the supercritical case. The large $u_{e\perp}$ and $u_{i\perp}$ in $2150 < \xi < 2750$ and $2920 < \xi < 3520$ are due to the transverse fluctuation of ions and electrons sustaining the Alfvén waves. Notice that the critical values $b_{w,1}$ and $b_{w,2}$ depend slightly on the initial parameter $b_w$. The number of particles per cell is 64 for both electrons and ions. The grid size is fixed to $\Delta x = \sqrt{2}/20 \times c'(=1)/\omega'_{pe}$. The initial temperatures of electrons and ions are $T_e = T_i = 1/1600 \times m'_e(=1)c'^2$. The simulation domain is in the range of $\xi = [0, 5670]$, where the periodic boundary condition is set for both particles and electromagnetic fields. The time step is determined as $\Delta t = 0.9 \Delta x$ to satisfy the Courant–Friedrichs–Lewy condition, and the total simulation time is $\tau = 2389$. The two counterpropagating waves start to collide at the center of the simulation domain ($\xi = 2835$) at $\tau_{\rm start} = 86$. The boundaries are located sufficiently far from the center so that simulation ends before the waves reach the boundaries.

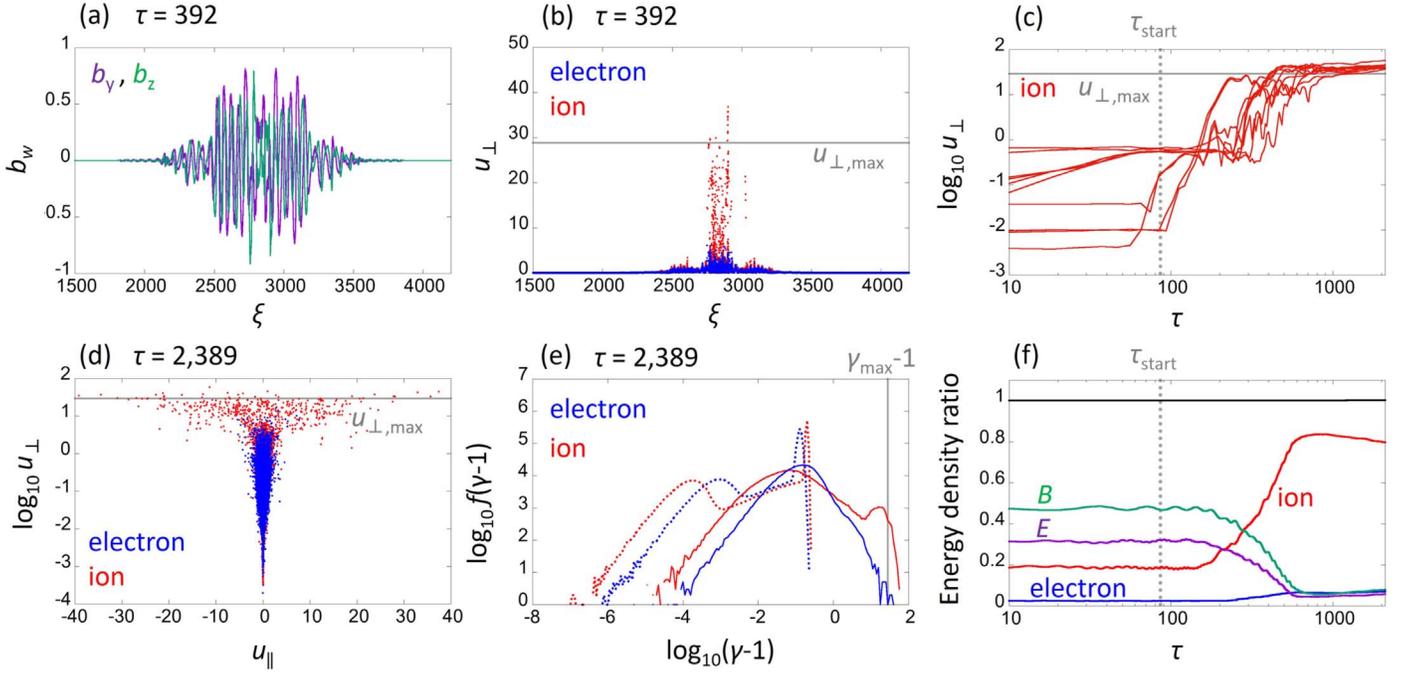

**Figure 8.** Results when the wave amplitude is supercritical $b_w = 0.6$ ($> b_{w,2}$). (a) $b_y$ (purple) and $b_z$ (green) profiles at $\tau = 392$. (b) Electron and ion distributions in $\xi$–$u_\parallel$ space at $\tau = 392$ (gray line denotes $u_{\perp \max} = 31$ estimated from Equation (10)). (c) Time histories of $u_\perp$ of the top 10 highest-energy ions at $\tau = \tau_{\rm end}$ (gray horizontal line and gray vertical dotted line denote $u_{\perp \max}$ and $\tau = \tau_{\rm start}$, respectively). (d) Electron and ion distributions in $u_\perp$–$u_\parallel$ space at $\tau = \tau_{\rm end}$ (gray line denotes $u_{\perp \max}$). (e) Energy distributions of electrons and ions at $\tau = \tau_{\rm end}$ (evaluated at $1500 < \xi < 4170$) (dotted lines indicate the initial distributions, and gray line denotes $\gamma_{\max} - 1$, where $\gamma_{\max} = \sqrt{1 + u_{\perp \max}^2}$). (f) Time histories of the energy density ratio of electromagnetic (EM) field (green), electron (blue), ion (red) and total energy (black).

When the wave amplitude is supercritical $b_w = 0.6 > b_{w,2}$, ions are suddenly accelerated to $u_{\perp \max}$ within $\tau \sim$ a few $\times 10^2$ (Figure 8(c)) after the two waves collide at the center. A large number of ions are accelerated to relativistic energy during the interaction of the two waves (Figures 8(a) and (b)). At the end of the interaction ($\tau = 2389$), as seen in Figure 8(e), the bulk of ions (and electrons) is accelerated. Over 70% of the total energy is converted to the ion's energy as shown in Figure 8(f). The number and the maximum energy of accelerated ions in Figure 8(e) are smaller than those in the test particle simulation (Figure 4(a)). Reduction of the maximum energy results from smaller $u_\parallel$ in Figure 8(d) than that in the test particle simulation (Figure 4(b)). These appear to be due to the back-reaction of the particle acceleration. When the wave amplitude is subcritical $b_w = 0.3 < b_{w,1}$, as seen in Figure 9, the two waves just pass through without significantly accelerating ions. The wave energy is converted to the electron's and ion's thermal energy.

Figure 10 shows the fraction of ion energy normalized by the injected wave energy at $\tau = 2389$. When the wave amplitude is supercritical, the fraction of ion energy reaches over 70% and increases with the wave amplitude up to 83% at $b_w = 0.8$, while it is less than 35% when the wave amplitude is subcritical.

## 4. Summary and Discussions

In summary we studied an efficient particle acceleration process in two counterpropagating circularly polarized Alfvén waves. By focusing on the motion of particles trapped in a trough of the magnetic envelope, we found that phase transition in the particle trajectory occurs when the amplitude of the wave magnetic field exceeds two critical values, $b_{w,1}$ and $b_{w,2}$. Below $b_{w,1}(< b_{w,2})$, the particles having initially nonrelativistic energy are forbidden to enter the relativistic trapping region in $u_\perp$–$\psi$ space. However, above $b_{w,1}$, nonrelativistic particles begin to be accelerated to relativistic energy, and above $b_{w,2}$, any nonrelativistic particles are accelerated to relativistic energy.

Since the acceleration time is very short, only a gyroperiod, it is enough to accelerate particles if the counterpropagating Alfvén waves are generated only shortly and locally in situations such as the successive decay instability of Alfvén waves. The counterpropagating situation can occur in various high-energy astrophysical phenomena. As mentioned in Section 3.1, this acceleration mechanism works even when the parameters of two waves are different. In the terrestrial radiation belt, whistler waves propagating back and forth along a dipole magnetic field are often observed. The collision between the interplanetary shock wave and the magnetospheric shock wave is also an example. The situation in which the waves generated in the solar flare loop top propagate down to the solar surface and are reflected may be another example. In the model of solar wind acceleration, it is often presumed that turbulence is driven by the collision between counterpropagating Alfvén waves (Shoda et al. 2019; Shestov et al. 2022). There is also a model in which a starquake can be a driver of bursty emission from a magnetar in which the energy of twisting magnetic field on the star surface is released as Alfvén waves (Israel et al. 2005; Wang 2020). If the Alfvén waves propagate along a closed magnetic field line, they may be reflected at the conjugate point of the star surface so that the counterpropagating situation is expected to be easily realized.

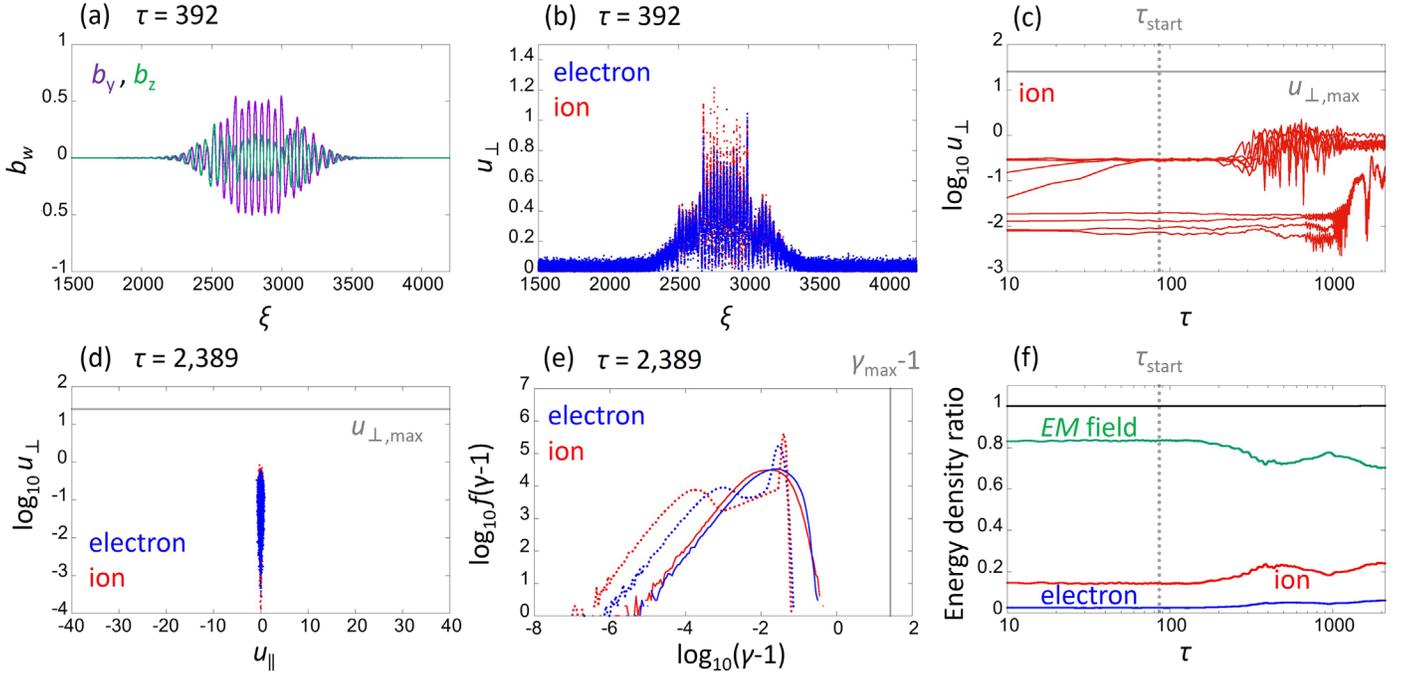

**Figure 9.** The same results when the wave amplitude is subcritical $b_w = 0.3$ ($<b_{w,1}$) as in Figure 6.

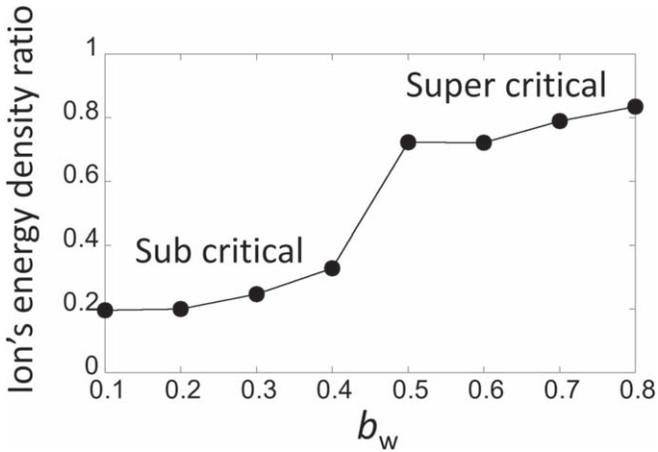

**Figure 10.** The dependence of the ion's energy density ratio obtained at $\tau = 2389$ on the wave amplitude where $b_w$ is supercritical at $b_w \geqslant 0.5$.

It is easily confirmed by assuming $\epsilon_s = -1$ and $\nu > 0$ that the same acceleration mechanism works also for electrons. Recently, it was reported using PIC simulation that efficient electron acceleration occurs in the interaction of a dense plasma foil with an intense laser under a strong external magnetic field (Sano et al. 2017, 2019, 2020). When electron cyclotron frequency is higher than the laser frequency, the acceleration is enhanced locally at the surface of the foil, where the incident laser is partly reflected. For incident lasers with wavelength $\sim 1\ \mu$m, the electron acceleration process will be demonstrated by applying more than 10 kT of an ambient magnetic field.

We confirmed that the acceleration occurs without assuming $u_\parallel = 0$ by performing test particle simulations. Furthermore, when $u_\parallel \neq 0$, all particles gain relativistic energy irreversibly within a short time regardless of their initial energy and position. Even when the back-reaction of the particle acceleration is taken into account in PIC simulation, a large number of ions are accelerated to relativistic energy before the waves are damped. The wave energy is efficiently converted to ions' kinetic energy.

Fast dissipation of counterpropagating linearly polarized Alfvén wave was also investigated in a strongly magnetized electron–positron plasma by Li et al. (2021). They studied a limiting case of long wavelength and low frequency with the Alfvén wave packet width of $\lambda = 3072 c/\omega_{pe}$. They showed that fast dissipation of two Alfvén waves is activated after two waves collide when the wave amplitude of each wave exceeds half of the background magnetic field ($b_w > 1/2$) and the magnetohydrodynamic condition $E < B$ breaks in the overlapping region of the two waves. The critical wave amplitude $b_w = 1/2$ is consistent with that given by Equations (13) and (15) in a long-wavelength and low-frequency limit. On the other hand, in this limit, the maximum attainable energy of a particle approaches infinity in Equation (11) and becomes $\gamma_{\max} = \sqrt{1 + u_{\perp\,\max}^2} = 16321$ when the same parameters of $b_w = 1.5$ and $\kappa = \nu = 4.6 \times 10^{-4}$ are used as in Li et al. (2021), while the averaged particle's energy gain is limited and its Lorenz factor is expected to be $\gamma_{\text{average}} = 82$ in Li et al. (2021). There may be some similarity and difference between their case and ours. To clarify the mutual relationship, further studies are necessary in the future.

This work was performed under the joint research project of the Institute of Laser Engineering, Osaka University. This research was supported by JSPS KAKENHI grant Nos. JP22K14020, JP21K03500, and JP20H00140 and the JSPS Core-to-Core Program, B. Asia-Africa Science Platforms No. JPJSCCB20190003. S.M. was supported by MEXT | Japan Society for the Promotion of Science (JSPS): JP22H01287.

### ORCID iDs

S. Isayama https://orcid.org/0000-0002-7531-8211
S. Matsukiyo https://orcid.org/0000-0002-4784-0301
T. Sano https://orcid.org/0000-0001-9106-3856

### References

Bellan, P. M. 2013, PhPl, 20, 042117
Blaes, O., Blandford, R., Goldreich, P., & Madau, P. 1989, ApJ, 343, 839
Blandford, R., & Eichler, D. 1987, PhR, 154, 1
Chen, P., Tajima, T., & Takahashi, Y. 2002, PhRvL, 89, 161101
Comisso, L., & Sironi, L. 2018, PhRvL, 121, 255101
Ebisuzaki, T., & Tajima, T. 2014, EPJST, 223, 1113
Ebisuzaki, T., & Tajima, T. 2021, APh, 128, 102567
Giacalone, J., Schwartz, S. J., & Burgess, D. 1993, GeoRL, 20, 149
Goldstein, M. L., Roberts, D. A., & Matthaeus, W. H. 1995, ARA&A, 33, 283
Hess, V. F. 1912, PhyZ, 13, 1084
Holcomb, K. A., & Tajima, T. 1991, ApJ, 378, 682
Hollweg, J. V., Esser, R., & Jayanti, V. 1993, JGR, 98, 3491
Israel, G. L., Belloni, T., Stella, L., et al. 2005, ApJL, 628, L53
Li, X., Beloborodov, A. M., & Sironi, L. 2021, ApJ, 915, 101
Ligorini, A., Niemiec, J., Kobzar, O., et al. 2021, MNRAS, 502, 5065
Matsukiyo, S., & Hada, T. 2003, PhRvE, 67, 046406
Matsukiyo, S., & Hada, T. 2009, ApJ, 692, 1004
Nariyuki, Y., & Hada, T. 2006, NPGeo, 13, 425
Sano, T., Fujioka, S., Mori, Y., Mima, K., & Sentoku, Y. 2020, PhRvE, 101, 013206
Sano, T., Hata, M., Kawahito, D., Mima, K., & Sentoku, Y. 2019, PhRvE, 100, 053205
Sano, T., Tanaka, Y., Iwata, N., et al. 2017, PhRvE, 96, 043209
Shestov, S. V., Voitenko, Y. M., & Zhukov, A. N. 2022, A&A, 661, A93
Shoda, M., Suzuki, T. K., Asgari-Targhi, M., & Yokoyama, T. 2019, ApJL, 880, L2
Terasawa, T., Hoshino, M., Sakai, J.-I., & Hada, T. 1986, JGR, 91, 4171
Wang, J.-S. 2020, ApJ, 900, 172